# Causal Impact Of European Union Emission Trading Scheme On Firm Behaviour And Economic Performance: A Study Of German Manufacturing Firms

*Nitish Gupta, Jay Shah, Satwik Gupta, Ruchir Kaul*

*Indian Institute of Technology Madras*

# 1 INTRODUCTION

## 1.1 FRAMEWORK

As a strong initiative to cost-effectively curb GreenHouse Gas (GHG) emissions from industrial installations and share the climate change burden, the EU came up with their Emissions Trading Scheme in 2005. It was different from traditional quota-based approach, in the sense that it was a "cap and trade" approach where polluters can trade the right to pollute among each other and thereby allow market forces equalize marginal abatement costs.

The EU Emissions Trading Scheme (EU ETS) covers more than 2 billion tons of $CO_2$ in 31 countries, making it the world's largest cap-and-trade system till date.

## 1.2 CONVENTIONAL ECONOMIC VIEW

The EU ETS puts a price on greenhouse gas emissions from regulated installations and influences the production and investment decisions of regulated firms. Regulated firms have to comply with requirements regarding the monitoring, reporting and verification of emissions. Consequently, they might have to utilize capacities in order to be able to trade allowances or to identify and implement emission reduction measures.

**There are two major concerns regarding this policy:**

1. Firstly, the EU ETS creates disadvantages for regulated firms that are exposed to competition from outside the EU. In particular, firms from the manufacturing sector that sell their goods and services on global markets might be vulnerable due to additional cost imposed through the EU ETS. Hence the belief is that it would unfairly put the regulated firms in the EU at a disadvantage

2. Secondly, these activities might bind otherwise productive inputs and therefore might have a detrimental effect on a firms' overall efficiency.

## 1.3 COUNTER ARGUMENT (PORTER HYPOTHESIS)

By contrast, Porter (1991) and Porter and van der Linde (1995) propose that *well defined and flexible environmental regulation* could help to overcome obstacles that impede process and product innovation, such as behavioural barriers, market failures others than the environmental externality, as well as organizational failure and thus might subsequently enhance firm performance.

All these aspects related to competitiveness and economic efficiency of regulated firms have been explored in detail and empirical evidence shows results **contrary to the conventional views.**

# 2 OBJECTIVES

We have two clear objectives:

1. *To estimate the causal impact (i.e. Average Treatment Effect, ATT) of the EU ETS on GHG emissions and firm competitiveness (primarily measured by employment, turnover, and exports levels) by combining a **difference-in-differences approach with semi-parametric matching techniques and estimators***

2. *To investigate the effect of the EU ETS on the economic performance of these German manufacturing firms using a **Stochastic Production Frontier model.***

*This is a novel approach since we use the firm specific distance to the frontier as a measure of economic performance. In other words, we relate input use and produced output and assess performance relative to the most efficient firms in the industry.*

The empirical models, methodology, applied statistics and inferences, and conclusions meeting both the objectives have been discussed.

# 3 DATA ANALYSIS FRAMEWORK

## 3.1 TIME FRAME

The window of our analysis covers two pretreatment years (2003–2004), phase I of the EU ETS, which ran from 2005 until 2007 and the initial years of phase II (2008-2012) . Phase I was completely decoupled from Phase II.

*Banking and borrowing was allowed across years within each compliance period, but not between Phase I and II.

## 3.2 DATA SOURCES

The principal dataset is the "AFiD-Betriebspanel" from Germany (which was made available to approved researchers at the Research Data Centres maintained by the German Federal Statistical Office and the statistical offices of the German Land, and whose published data from an academic paper we used).This data serves as a basis for many official German governmental statistics and includes all manufacturing firms with more than 20 employees.

It covers approximately 50,000 plants per year and contains information on a wide range of economic variables such as employment, gross output, investment, and exports. These data are collected as part of the monthly production surveys administered by the German statistical office.

# 4 COUNTRY AND INDUSTRY SELECTION

## 4.1 WHY GERMANY?

Our study focuses on Germany, Europe's largest economy and also its largest emitter. More than 1,900 of all EU ETS installations are based in Germany, accounting for approximately one fifth of total regulated $CO_2$ emissions. Also, the fact that most of the German manufacturing sector is more oriented towards export markets, Germany is a particularly interesting case for investigating the validity of the first major concern on this proposal, that the unilateral regulation of EU firms leads to a loss of competitiveness of these firms in the international markets.

## 4.2 WHICH MANUFACTURING SECTORS?

According to the Emissions Trading Directive, participation in the EU ETS is mandatory for all combustion installations with a rated thermal input of 20 MW or more and certain process specific industries operating above the predetermined capacity thresholds. This concerns *mostly heat and power generation*.

Our focus would be on four major industries, namely:

1. "Manufacture of paper and paper products" (WZ classification code 17);
2. "Manufacture of coke and refined petroleum products" (WZ19);
3. "Manufacture of other non-metallic mineral products" (WZ23, including i.a. manufacture of glass, ceramics, and cement),
4. and "Manufacture of basic metals" (WZ24).

## 5 DESIGN OF THE RESEARCH MODEL 1

### 5.1 EMPIRICAL FRAMEWORK

In line with the potential outcome framework,

We denote by $Y_i(1)$ the outcome at firm i when subject to the ETS and by $Y_i(0)$ the outcome when the plant is not subject to the ETS.

Let $D_i$ denote the treatment indicator, and subscripts t' and t denote pre- and post-treatment periods, respectively. X is a set of observable covariates.

We are interested in the average treatment effect on the treated (ATT) i.e.

$$\alpha_{ATT} = E\left(Y_{it}(1) - Y_{it}(0) | X, D = 1\right).$$

### 5.2 SELECTION BIAS AND CONFOUNDEDNESS PROBLEM TO ISOLATE THE CAUSAL EFFECT OF EU ETS:

The fundamental evaluation problem arises because $Y_{it}(0)$ is unobserved for the treated.

To get over the selection bias of treated firms due to variations in technical characteristics such as production capacity, process regulation, etc amongst themselves and control firms, we use the matching approach which consists of imputing $Y_{it}(0)$ using outcomes for untreated firms that are observationally equivalent to the treated firm.

To mathematically implement this, we propose a semiparametric conditional DD matching estimator

$$\hat{\alpha} = \frac{1}{N_1} \sum_{i \in I_1} \left\{ (Y_{it}(1) - Y_{i0}(0)) - \sum_{k \in I_0} W_{N_0, N_1}(i, k) \cdot (Y_{kt}(0) - Y_{k0}(0)) \right\}$$

where $I_1$ is the set of $N_1$ ETS participants and $I_0$ is the set of $N_0$ non-participants. The weight $W_{N_0,N_1(i,k)}$ with $P_k \in I_0$. $W_{N_0,N_1(i,k)} = 1$ determines how strongly the counterfactual observation k contributes to the estimated treatment effect. For instance, a control plant is weighted more strongly the more similar – in covariate space – it is to the treated facility.

**In summary, this selection bias was tackled using:**

1. Weighing the control firms and then pairing treated and untreated firms having similar propensity scores using nearest neighbour matching (NN) techniques ( 1:1 till 1:20 for robustness checks.

2. ATT estimation using a combination of weighting and regression Specifically, we perform OLS on the weighted DD
equation

$$\Delta y_{it} = constant + \alpha^R_{ATT} D_i + x'_{it}\beta + \epsilon_{it}$$

where the propensity score is used to reweight the distribution of treated and non-treated firms.

Contrary to NN matching, weighted outcomes of all untreated firms are used to construct the counterfactual. While treated firms enter the regression with a weight of one, the weights for the untreated firms pk 1-pk are based on the estimated propensity score and ensure that the distribution of the control variables is approximately equal for both groups.

## 5.3 ASSUMPTIONS

1. Counterfactual trends in outcomes in ETS firms must not be systematically different from those in the group of matched control firms.
2. Matching is performed on those outcome covariates whose distribution overlaps in both groups.
3. No Spillover between the treated and control firms ( SUTVA Stable Unit Treatment Value Assumption Assumption)

## 6 METHODOLOGY

1. Identification of outcome variables of interest for both the full sample and by treatment status for the year 2003, i.e., the last pretreatment year

2. Correcting for confoundedness and bias using
Matching techniques (such as one-to-one NN) by comparing propensity scores
OLS probit regression with propensity reweighting by analysing the equality and trends in the pre-treatment year.

3. Constructing counterfactual for each treated firm and analysing the average treatment effect on the treated-on outcome covariates of interest. Since energy use is a central aspect of this study information on energy consumption for more thadifferent fuel types, electricity generation on site and electricity trading allows us to accurately calculate plant-level carbon emissions and carbon intensity of production.

# 7 Descriptive Statistics And Matching Techniques using pre-treatment years (2002-2003)

Table 1: Summary statistics for outcome variables and covariates in 2003

| Variable | (1) Mean | (2) Std. Dev. | (3) p10 | (4) p50 | (5) p90 | (6) N |
|---|---|---|---|---|---|---|
| **A. Full sample (mid-98%)** | | | | | | |
| $CO_2$ emissions from energy (t) | 1,912 | 5,618 | 35 | 314 | 4,098 | 40,834 |
| $CO_2$ intensity (g/€1000) | 108,581 | 143,612 | 8,250 | 62,793 | 248,907 | 40,709 |
| Employees | 104 | 158 | 22 | 49 | 233 | 40,325 |
| Gross output (€1000) | 17,597 | 38,223 | 1,435 | 5,299 | 40,580 | 40,204 |
| Exports (€1000) | 4,978 | 15,776 | 0 | 198 | 11,542 | 40,947 |
| Export share of output | 0.16 | 0.22 | 0.00 | 0.04 | 0.53 | 40,931 |
| Average wage rate (€) | 28,649 | 9,681 | 15,998 | 28,458 | 41,213 | 40,409 |
| **B. ETS participants** | | | | | | |
| $CO_2$ emissions from energy (t) | . | 795,888 | 6,146 | 51,716 | 457,851 | 408 |
| $CO_2$ intensity (g/€1000) | . | 1,449,921 | 84,392 | 670,420 | 2,604,891 | 413 |
| Employees | . | 11,370 | 52 | 388 | 4,103 | 433 |
| Gross output (€1000) | . | 4,191,998 | 6,748 | 95,703 | 1,125,042 | 430 |
| Exports (€1000) | . | 2,853,722 | 324 | 28,064 | 647,477 | 369 |
| Export share of output | . | 0.27 | 0.00 | 0.29 | 0.70 | 408 |
| Average wage rate (€) | . | 9,393 | 26,729 | 37,214 | 48,408 | 408 |
| **C. Non-ETS participants (matched sample)** | | | | | | |
| $CO_2$ emissions from energy (t) | . | 372,759 | 510 | 12,047 | 891,534 | 278 |
| $CO_2$ intensity (g/€1000) | . | 1,786,216 | 37,991 | 155,349 | 1,769,886 | 283 |
| Employees | . | 1,994 | 42 | 208 | 4,384 | 296 |
| Gross output (€1000) | . | 759,593 | 5,728 | 64,809 | 825,606 | 293 |
| Exports (€1000) | . | 323,759 | 936 | 21,537 | 802,049 | 248 |
| Export share of output | . | 0.25 | 0.00 | 0.28 | 0.64 | 278 |
| Average wage rate (€) | . | 9,283 | 25,767 | 39,210 | 49,010 | 278 |

Notes: $CO_2$ intensity in terms of gross output (g/€1000). Means for matched sample cannot be obtained for reasons of data privacy.
Source: Research Data Centres of the Federal Statistical Office and the Statistical Offices of the Länder (2012): AFiD-Panel Industriebetriebe, 2005-2010, own calculations.

Upon comparing panels A and B of **Table 1**, we notice that approximately 1 percent of our full sample consists of ETS participants. Moreover, ETS participants are considerably larger, more prone to export, and better paying throughout the distribution of firms, as well as more heterogeneous than the full sample of firms. This highlights the extent of selection on observable firm characteristics.

Panel C reports the final descriptive statistics. The systematic differences between ETS firms and non-ETS firms vanish after matching, compared to the differences of the full sample. Notice that the 1st decile is smaller for the non-ETS participants compared to ETS participants in most cases, just as the 9th decile is usually larger. This justifies our assumption that the support of the distribution of ETS participants is overlapped by the distribution of non-ETS participants for most variables.

Table 2: Pre-treatment outcomes in the matched sample

| | Null hypothesis: Equality of pre-treatment outcomes | | | | | |
|---|---|---|---|---|---|---|
| | A. Levels | | | B. Trends | | |
| | | Number of | | | Number of | |
| Variable | p-value | treated | controls | p-value | treated | controls |
| $CO_2$ emissions | 0.0911 | 408 | 278 | 0.0505 | 405 | . |
| $CO_2$ intensity | 0.0197 | 413 | 283 | 0.2025 | 409 | . |
| Gross output | 0.0054 | 430 | 293 | 0.3141 | 428 | . |
| Employees | 0.0051 | 433 | 296 | 0.6177 | 431 | . |
| Exports | 0.0073 | 369 | 248 | 0.1047 | 336 | . |
| Export share | 0.1634 | 408 | 278 | 0.2483 | 406 | . |
| Average wage rate | 0.0086 | 408 | 278 | 0.0603 | 285 | . |

Notes: Number of control firms for matched sample are not reported for confidentiality reasons.
Source: Research Data Centres of the Federal Statistical Office and the Statistical Offices of the Länder (2012): AFiD-Panel Industriebetriebe, 2005-2010, own calculations.

Panel A of **Table 2** reports the results of a test of equality of means in pre-treatment levels of ETS participants and non-participants. To construct a meaningful comparison group for the ETS participants, we select a limited number of non-ETS participants by means of semi-parametric matching.

Apart from equality of pre-treatment levels across groups, we test for equality of pre-treatment trends in panel B, based on logged differences between 2002 and 2003. Equality of pre-treatment trends is not rejected for the matched sample, with p-values above 5 percent for all outcome variables.

This sheds some light on the extent to which the common trends assumption underlying a DID matched estimator is substantiated by the dataset.

# 8 Main Results

The principal objective of the EU ETS is to reduce carbon dioxide emissions in Europe. Estimating the ATT of the EU ETS on emissions of treated firms tells us how successful the policy has been at curbing CO2 emissions in the German manufacturing sector.

**1)Table 3 displays the ATT estimates for CO2 emissions**.

Columns (1) and (2) report coefficients estimated with one-to-one and one-to-twenty nearest-neighbour matching, respectively. Column (3) reports the coefficient estimated using the reweighted OLS estimator from (4). Columns (4) and (5) report numbers of treated and control observations for nearest neighbour matching.

Panel A reports the log change in CO2 emissions that can be causally attributed to participation in the EU ETS, whereas panel B reports the causal effect in terms of the log change in the carbon intensity of output. The estimates are reported separately for phase I (2005-2007) and the first half of phase II (2008-2010).

Table 3: Impact on $CO_2$ emissions

|  | (1) | (2) | (3) | (4) | (5) |
|---|---|---|---|---|---|
|  | \multicolumn{3}{c}{Estimation Algorithm} | \multicolumn{2}{c}{Number of} |
|  | NN (1:1) | NN (1:20) | OLS w/R | Treated | Controls |
| **A. $CO_2$ emissions: $\Delta \ln(CO_2)$** | | | | | |
| Phase I | 0.00 | 0.02 | 0.03 | 452 | 27,710 |
|  | (0.03) | (0.02) | (0.03) | | |
| Phase II | -0.28** | -0.25** | -0.26** | 408 | 23,908 |
|  | (0.05) | (0.03) | (0.03) | | |
| **B. $CO_2$ intensity of gross output: $\Delta \ln(\frac{CO_2}{GO})$** | | | | | |
| Phase I | 0.04 | 0.03 | 0.05* | 451 | 27,637 |
|  | (0.05) | (0.03) | (0.03) | | |
| Phase II | -0.18** | -0.20** | -0.30** | 412 | 23,742 |
|  | (0.05) | (0.04) | (0.03) | | |

Notes: NN(1:1) and NN(1:20) denote nearest neighbor matching with one and 20 neighbors, respectively. OLS w/R denotes the reweighted OLS estimator. Standard errors in parenthesis. *** p<0.01, ** p<0.05, * p<0.1. Source: Research Data Centres of the Federal Statistical Office and the Statistical Offices of the Länder (2012): AFiD-Panel Industriebetriebe, 1998-2010, own calculations.

**Inference:**

A clear pattern emerges from panel A. The point estimates for the first trading phase are positive, very close to zero and lack statistical significance. We thus cannot reject the Null hypothesis that treated firms conducted no abatement in the first phase. In contrast, we see strong evidence that phase II of the EU ETS caused treated firms to reduce their emissions by a substantial margin, in the order of 25 to 28 percentage points more than non-treated firms. This finding is statistically significant at the 5% level and robust across specifications.

The impact on carbon intensity, reported in Panel B, closely mimics the overall effect on emissions. Carbon intensity remains almost unchanged throughout phase I, although the point estimates are somewhat larger than for emissions and the reweighting estimator actually yields an increase by 5 percentage points at the 10 percent significance level. However, in phase II, carbon intensity fell between 18 and 30 percentage points faster at EU ETS firms than at the control firms, and again this effect is statistically significant at the 5 percent level.

*This result suggests that firms responded to the introduction of the EU ETS mainly by adjusting intensity, not scale of production.*

**In order to justify our result that the production levels of firms remain unaffected we examine the changes in output and employment in Table 4.**

Table 4: Impact on employment, revenue and exports

|  | (1) | (2) | (3) | (4) | (5) |
|---|---|---|---|---|---|
|  | Estimation Algorithm | | | Number of | |
|  | NN (1:1) | NN (1:20) | OLS w/R | Treated | Controls |
| A. Employees ($\Delta \ln L$) | | | | | |
| Phase I | 0.00 | -0.02 | -0.02 | 454 | 28,396 |
|  | (0.02) | (0.01) | (0.01) | | |
| Phase II | 0.03 | 0.01 | 0.01 | 433 | 24,237 |
|  | (0.02) | (0.01) | (0.01) | | |
| B. Gross output ($\Delta \ln GO$) | | | | | |
| Phase I | 0.01 | 0.01 | 0.01 | 449 | 28,465 |
|  | (0.03) | (0.02) | (0.02) | | |
| Phase II | 0.07*** | 0.05*** | 0.04** | 430 | 24,240 |
|  | (0.03) | (0.02) | (0.02) | | |
| C. Exports: $\Delta \ln(X)$ | | | | | |
| Phase I | 0.06 | 0.10** | 0.11*** | 371 | 17,864 |
|  | (0.06) | (0.04) | (0.04) | | |
| Phase II | 0.18*** | 0.09** | 0.07* | 348 | 15,463 |
|  | (0.06) | (0.04) | (0.04) | | |

Notes: NN(1:1) and NN(1:20) denote nearest neighbor matching with one and 20 neighbors, respectively. OLS w/R denotes the reweighted OLS estimator. *** $p < 0.01$, ** $p < 0.05$, * $p < 0.1$.
Source: Research Data Centres of the Federal Statistical Office and the Statistical Offices of the Länder (2012): AFiD-Panel Industriebetriebe, 1998-2010, own calculations.

Panel A of **Table 4** reports the average treatment effects of the EU ETS on employment. The point estimates for employment range from -0.02 to 0.01 log points. None of the coefficient estimates is statistically significant at the 5% level.

*Hence our results do not support fears that putting a price on carbon comes at the expense of domestic job destruction.*

The estimated impact of the EU ETS on gross output, reported in Panel B is small and insignificant in phase I. In phase II, however, we estimate that the EU ETS increased gross output at regulated firms by a statistically significant amount of between 4 and 7 percent.

*We thus reject the hypothesis that the EU ETS caused firms to reduce the scale of production. Infact, the gross output has increased suggesting that the ETS made those firms adapt policies and use technologies that improved/maintained productivity, while also substantially reducing emissions.*

The positive effect on gross output is consistent with both firms producing more and charging higher prices.

Although it isn't clear whether an increase in exports reflects an increase in shipments or prices, but we can certainly reject the hypothesis that the EU ETS caused regulated firms to reduce their overall exports.

# 9 So, how did treated firms reduce carbon emissions?

A robust finding established in the previous section is that the EU ETS caused firms to substantially cut back on CO2 emissions in phase II. Further, this cutback was achieved through a reduction in the carbon intensity rather than the scale of production. To probe deeper, we draw on additional data from various sources to shed more light on how treated firms reduced the carbon footprint of production.

From a conceptual point of view, firms can achieve this by reducing the carbon intensity for a given level of energy consumption – for example by switching from high-carbon fuels to low-carbon fuels – or by using energy more efficiently for a given energy mix.

## 9.1 Evidence for Fuel switching

Table 6: Impact on fuel use

|  | (1) | (2) | (3) | (4) | (5) |
|---|---|---|---|---|---|
|  | \multicolumn{3}{c}{Estimation Algorithm} | \multicolumn{2}{c}{Number of} |
|  | NN (1:1) | NN (1:20) | OLS w/R | Treated | Controls |
| **A. Electricity consumption: $\Delta \ln(ELEC)$** | | | | | |
| Phase I | 0.01 | 0.03 | 0.02 | 453 | 27,699 |
|  | (0.03) | (0.02) | (0.03) |  |  |
| Phase II | -0.04 | -0.03 | -0.04** | 428 | 23,867 |
|  | (0.03) | (0.02) | (0.02) |  |  |
| **B. Consumption of all non-electricity fuels: $\Delta \ln(EPRIMARY)$** | | | | | |
| Phase I | 0.16*** | 0.13*** | 0.11*** | 435 | 24,601 |
|  | (0.05) | (0.04) | (0.03) |  |  |
| Phase II | -0.81** | -0.83** | -0.87** | 376 | 21,331 |
|  | (0.15) | (0.11) | (0.1) |  |  |
| **C. Consumption of natural gas: $\Delta \ln(GAS)$** | | | | | |
| Phase I | 0.01 | 0.10** | 0.11** | 412 | 16,817 |
|  | (0.07) | (0.04) | (0.04) |  |  |
| Phase II | -0.21** | -0.32** | -0.33** | 217 | 10,506 |
|  | (0.11) | (0.08) | (0.07) |  |  |
| **D. Consumption of petroleum products: $\Delta \ln(OIL)$** | | | | | |
| Phase I | -0.05 | -0.02 | -0.15** | 232 | 8,857 |
|  | (0.12) | (0.08) | (0.07) |  |  |
| Phase II | -0.56** | -0.45** | -0.48** | 163 | 7,815 |
|  | (0.17) | (0.11) | (0.13) |  |  |

Notes: NN(1:1) and NN(1:20) denote nearest neighbor matching with one and 20 neighbors, respectively. OLS w/R denotes the reweighted OLS estimator. Standard errors in parenthesis. *** p<0.01, ** p<0.05, * p<0.1. Source: Research Data Centres of the Federal Statistical Office and the Statistical Offices of the Länder (2012): AFiD-Panel Industriebetriebe, 1998-2010, own calculations.

**In Table 6**, Panel A shows that ATT estimates for electricity consumption are between -0.03 and -0.04 in phase II, but only the OLS reweighting estimate is also statistically significant. In contrast, the estimated ATTs for non-electricity fuels, reported in Panel B, are highly statistically significant, with an increase of between 0.11 and 0.16 log points in phase I and a subsequent decrease by 0.81 to 0.87 log points in phase II. As Panels C and D show, this decrease is explained by the strong reductions in both natural gas and oil consumption, with point estimates between -0.21 to -0.33 for the former and between -0.45 and -0.56 for the latter. Moreover, the number of firms consuming natural gas and petroleum products falls by a larger proportion among treated than among untreated firms, as is evident from the last two columns. The point estimates for phase I suggest that some firms engaged in substitution of natural gas for oil, but this effect is not statistically significant in our preferred specification, and – as we already know – did not result in a significant reduction of overall carbon emissions

Overall, these findings suggest that treated firms pursued different strategies to cope with carbon pricing in the two trading phases. While treated firms first switched from high- to low-carbon content among non-electricity fuels, they drastically reduced their use of fossil fuels in phase II. It appears that carbon pricing in phase II increased the cost of generating heat or electricity on site beyond economically viable levels for the average treated firm. Less heat generation on site could mean that treated firms were making more efficient use of process heat, or that they shut down or throttled their on-site power plants.

### 9.2 Technology upgrades and other emissions reducing measures

To the extent that the substantial reduction in carbon emissions during phase II of the EU ETS cannot be attributed to the substitution toward fuels with a lower emissions intensity, it is likely the result of increased energy conservation efforts. For instance, the adoption of more efficient technologies tends to reduce energy use. This possibility was explored by using a "double-blind" telephone interview method from a broad-based survey of managers at medium-sized European manufacturing firms.

Table 8: Adoption of emissions reducing measures

|  | (1) All measures adopted | (2) | (3) Most significant measure | (4) |
|---|---|---|---|---|
|  | Share of adopters (%) | Effect of ETS | Share of adopters (%) | Effect of ETS |
| **I. Heating and cooling** | | | | |
| 1. Optimized use of process heat | 37.7*** (4.1) | 1.02*** (0.31) | 20.5*** (3.8) | 1.01*** (0.34) |
| 2. Modernization of cooling / refrigeration system | 9.4*** (2.5) | -0.22 (0.30) | 0.9 (0.9) | |
| 3. Optimization of air conditioning system | 4.4** (1.7) | 0.15 (0.42) | 0.9 (0.9) | |
| 4. Optimization of exhaust air system / district heating system | 27.5*** (3.8) | 0.01 (0.23) | 9.8*** (2.8) | -0.64* (0.34) |
| **II. More climate-friendly energy generation on site** | | | | |
| 1. Installation of CHP plant | 13.0*** (2.9) | 0.17 (0.28) | 6.3*** (2.3) | 0.05 (0.34) |
| 2. Biogas feed-in into local CHP plant or domestic gas grid | 2.9** (1.4) | 0.20 (0.46) | 5.4** (2.1) | |
| 3. Switching to natural gas | 2.9** (1.4) | -0.38 (0.44) | 0.9 (0.9) | |
| 4. Exploitation of renewable energy source | 13.8*** (2.9) | -0.17 (0.34) | 9.8*** (2.8) | 0.08 (0.50) |
| **III. Machinery** | | | | |
| 1. Modernization of compressed air system | 14.5*** (3.0) | 0.07 (0.25) | 5.4** (2.1) | -0.29 (0.43) |
| 2. Other industry-specific production process optimization/machine upgrade | 63.0*** (4.1) | 0.41** (0.20) | 23.2*** (4.0) | 0.29 (0.27) |
| 3. Production process innovation | 8.0*** (2.3) | 0.13 (0.37) | 1.8 (1.3) | |
| **IV. Energy management** | | | | |
| 1. Introduction of energy management system | 8.0*** (2.3) | -0.14 (0.30) | 1.8 (1.3) | |
| 2. Submetering / upgrade of existing energy management system | 7.3*** (2.2) | 0.56 (0.42) | 0.9 (0.9) | |
| 3. (External) energy audit | 7.3*** (2.2) | -0.14 (0.30) | - | |
| 4. Installation of timers attached to machinery | 4.4** (1.7) | -0.89** (0.35) | - | |
| 5. Installation of heating systems | 2.2* (1.3) | -0.59 (0.51) | 2.7* (1.5) | -0.52 (0.52) |
| **V. Other measures on production site** | | | | |
| 1. Modernization of lighting system | 12.3*** (2.8) | -0.68** (0.32) | 1.8 (1.3) | |
| 2. Energy-efficient site extension/ improved insulation/building management | 18.1*** (3.3) | -0.87*** (0.23) | 5.4** (2.1) | -0.76 (0.46) |
| 3. Employee awareness campaigns and staff trainings | 12.3*** (2.8) | -0.29 (0.28) | - | |
| 4. Non-technical reorganization of the production process | 2.2* (1.3) | -0.13 (0.41) | 0.9 (0.9) | |
| 5. Installation of energy efficient IT system | 6.5*** (2.1) | 0.10 (0.40) | - | |
| 6. Improved waste management / recycling | 5.1*** (1.9) | 0.54 (0.45) | 0.9 (0.9) | |

Notes: Based on telephone interviews with managers of 138 German manufacturing firms, 95 of which were EU ETS participants in 2009. Columns (1) and (3) report the mean and standard deviation (in parentheses) of the adoption rate for a given measure. Columns (2) and (4) report the coefficient on EU ETS participation in a probit regression of adoption, controlling for employment size, interviewer fixed effects, and respondent characteristics. Standard errors in parentheses are clustered at the 3-digit sector level.

# 10 CONCLUSION

## 10.1 REDUCTION IN CARBON EMISSIONS-

Our results indicate that the EU ETS did not reduce emissions in significant ways during its first phase, but it caused participating firms to substantially reduce their carbon emissions relative to untreated firms during phase II. ***This abatement was achieved through a reduction in the carbon intensity rather than through a reduced scale of production.***

While phase I saw some substitution of low-carbon for high-carbon fuels at treated firms, in phase II treated firms drastically reduced their use of all primary energy while maintaining constant their level of electricity consumption. We attribute this outcome to firms curbing onsite generation of heat and improved recovery of waste heat as the predominant way of reducing compliance costs. Qualitative evidence from telephone interviews with managers suggests that regulated firms optimized their use of process heat. In contrast, we find no evidence of major technological upgrades that would explain the reduction of carbon intensity.

## 10.2 AFFECT OF EU ETS ON FIRM COMPETITIVENESS.

Our second main result is that the EU ETS had no negative effect on gross output, employment or exports over the sample period.

# 11 EMPIRICAL STRATEGY AND METHODOLOGY FOR RESEARCH OBJECTIVE 2

## 11.1 Estimating the stochastic production frontier.

This function expresses the maximum amount of output that can be produced from a given set of inputs with a fixed technology given as:

$$\ln y_{it} = \ln f(\mathbf{x}_{it}) + \nu_{it} + u_{it}$$

1. $y_{it}$ denotes the output of firm i at year t
2. $f(x_{it})$ is the deterministic production frontier,
3. $x_{it}$ is a vector of inputs, including capital stock, labour and energy use
4. $\nu_{it}$ is a nonpositive random variable depicting inefficiency (drawn from a truncated normal distribution), and
5. $u_{it}$ is an independently and identically distributed error term with zero mean and constant variance (drawn from a symmetric normal distribution).

We assume the deterministic frontier $f(x_{it})$ to take the form of a Cobb-Douglas function and implement the model using <u>maximum likelihood estimation</u>.

In order to account for industry specific technologies, we estimate the stochastic frontier model for each two-digit industry within the German <u>manufacturing sector</u>. The distance to the frontier refers to a joint frontier for the years from 2003 to 2012.

The <u>estimated distance</u> to the frontier also captures <u>dynamic factors</u> that might drive firm's efficiency, such as <u>technological change</u>. Our identification strategy will take these characteristics of the distance to the production frontier into account

## 11.2 Identifying the effect of the EU ETS by estimating the sample average treatment effect on the treated (SATT)

We exploit this variation created by the inclusion criteria of the EU ETS in order to isolate the effect of the EU ETS on the distance between regulated firms and the efficient production frontier

We differentiate between treatment and control group depending on whether a firm has to comply with the regulation by the EU ETS or not.

Let the binary variable $ETS_i \in \{0, 1\}$ be an indicator that describes the treatment status of firm i. Let $ETS_i$ be equal 1 if the firm operates installations that are regulated by the EU ETS and 0 if the firm is not required to participate in the EU ETS.

Accordingly, we describe the potential outcomes by $Outcome_i(1)$ and $Outcome_i(0)$ for treatment and control group, respectively.

$$\tau = E[\text{Outcome}_{it}(1) - \text{Outcome}_{it}(0)|\text{ETS}_i = 1]$$

where τ is the average effect of the EU ETS on the distance between regulated firms and the efficient production frontier after the implementation of the EU ETS.

Similar to the previous model, while we are able to observe $Outcome_{it}(1)$ for regulated firms, the outcome $Outcome_{it}(0)$ is not realized in the case of regulated firms. Therefore, we will use information on the outcome $Outcome_{it}(0)$ collected from the firms that belong to the control group in order to form an adequate counterfactual. The comparison of the two groups will only lead to robust results, if factors that are correlated with efficiency dynamics do not differ across treatment and control group.

## 11.3 Nearest neighbour matching in order to account for potential bias.

The intuition behind this approach is to form a control group using unregulated firms that resemble the firms in the treatment group and thus might be affected by unobservable confounding factors in the same way.

Although, we do not pose any parametric assumptions on the relationship between the distance to the frontier and the explanatory variables $z_{it}$, however, we still rely on the conditional unconfoundedness and SUTVA where the common support assumption is of particular importance

## 12. DESCRIPTIVE STATISTICS

1) The table below depicts the variables used in the study for the entire manufacturing sector

Descriptive statistics German production census.

| | Mean | SD | Skewness | Kurtosis | P10 | P50 | P90 | N |
|---|---|---|---|---|---|---|---|---|
| **2003** | | | | | | | | |
| Output (in 1000 EUR) | 28,699.22 | 360,632.64 | 81.42 | 8262.81 | 1324.71 | 5276.71 | 42,903.50 | 37,888 |
| Emissions (in t CO2) | 7343.68 | 119,327.97 | 49.20 | 2996.19 | 64.39 | 374.07 | 5236.04 | 36,985 |
| Capitalstock (in 1000 EUR) | 11,322.21 | 120,382.53 | 53.92 | 3524.23 | 256.28 | 1838.52 | 16,220.64 | 37,099 |
| Number of employees | 153.06 | 1292.85 | 71.69 | 6278.85 | 22.58 | 50.00 | 254.00 | 38,319 |
| Energy use (in MWh) | 21,418.67 | 376,270.67 | 49.81 | 2991.01 | 161.79 | 911.97 | 12,979.47 | 36,949 |
| **2006** | | | | | | | | |
| Output (in 1000 EUR) | 33,903.31 | 417,111.75 | 77.80 | 7509.40 | 1483.30 | 6201.49 | 49,867.54 | 36,162 |
| Emissions (in t CO2) | 8978.83 | 193,026.98 | 69.25 | 6186.59 | 72.56 | 412.27 | 5742.81 | 35,654 |
| Capitalstock (in 1000 EUR) | 11,097.78 | 121,072.64 | 60.25 | 4516.43 | 248.68 | 1774.37 | 15,880.93 | 36,073 |
| Number of employees | 154.01 | 1313.92 | 73.58 | 6446.77 | 23.90 | 52.92 | 254.75 | 36,632 |
| Energy use (in MWh) | 27,200.64 | 618,927.15 | 64.61 | 5055.30 | 186.25 | 994.59 | 14,296.39 | 35,631 |
| **2009** | | | | | | | | |
| Output (in 1000 EUR) | 29,257.35 | 345,810.11 | 77.27 | 7309.36 | 1295.66 | 5346.61 | 44,534.20 | 36,703 |
| Emissions (in t CO2) | 7989.21 | 179,143.00 | 69.44 | 5857.38 | 66.90 | 362.56 | 5017.29 | 36,100 |
| Capitalstock (in 1000 EUR) | 11,148.64 | 119,355.75 | 60.18 | 4565.63 | 234.60 | 1785.08 | 16,464.80 | 36,335 |
| Number of employees | 152.47 | 1219.86 | 70.98 | 6127.83 | 24.00 | 53.00 | 254.50 | 36,982 |
| Energy use (in MWh) | 26,043.38 | 627,994.42 | 70.33 | 5934.58 | 179.39 | 920.20 | 13,337.29 | 36,074 |
| **2012** | | | | | | | | |
| Output (in 1000 EUR) | 35,194.48 | 514,872.08 | 89.04 | 9350.33 | 1431.54 | 6184.66 | 51,415.54 | 36,882 |
| Emissions (in t CO2) | 9012.12 | 211,455.53 | 71.14 | 6276.90 | 68.58 | 385.09 | 5489.93 | 36,435 |
| Capitalstock (in 1000 EUR) | 10,641.92 | 122,387.63 | 58.11 | 4256.77 | 240.51 | 1611.20 | 14,924.90 | 36,380 |
| Number of employees | 157.09 | 1300.76 | 69.85 | 5901.93 | 25.00 | 54.83 | 260.33 | 37,130 |
| Energy use (in MWh) | 29,383.97 | 745,139.74 | 73.12 | 6467.69 | 183.05 | 951.67 | 14,134.01 | 36,421 |

Notes: Output (production value) and capital stock are denoted in 1000 EUR. Energy use is denoted in MWh and $CO_2$ emissions in t $CO_2$ equivalent. Source: Research Data Centres of the Statistical Offices Germany (2014): Official Firm Data for Germany (AFiD) – AFiD-Panel Industrial Units and AFiD-Module Use of Energy, own calculations.

**Noteworthy observations are:**

1. The output as well as the use of inputs increase over time. However, the <u>economic crisis</u> led to declining output, emissions, and energy use in late 2008 and 2009 owing to decreasing orders.
2. The number of employees was not much affected e.g. due to governmental support programs and strict <u>labour market</u> regulation.
3. Although, the capital stock is quite stable, however, it slightly decreased in the aftermath of the crisis due to low investments during the crisis.

2) The following figure shows the development of the variables over time across two-digit industries within the manufacturing sector.

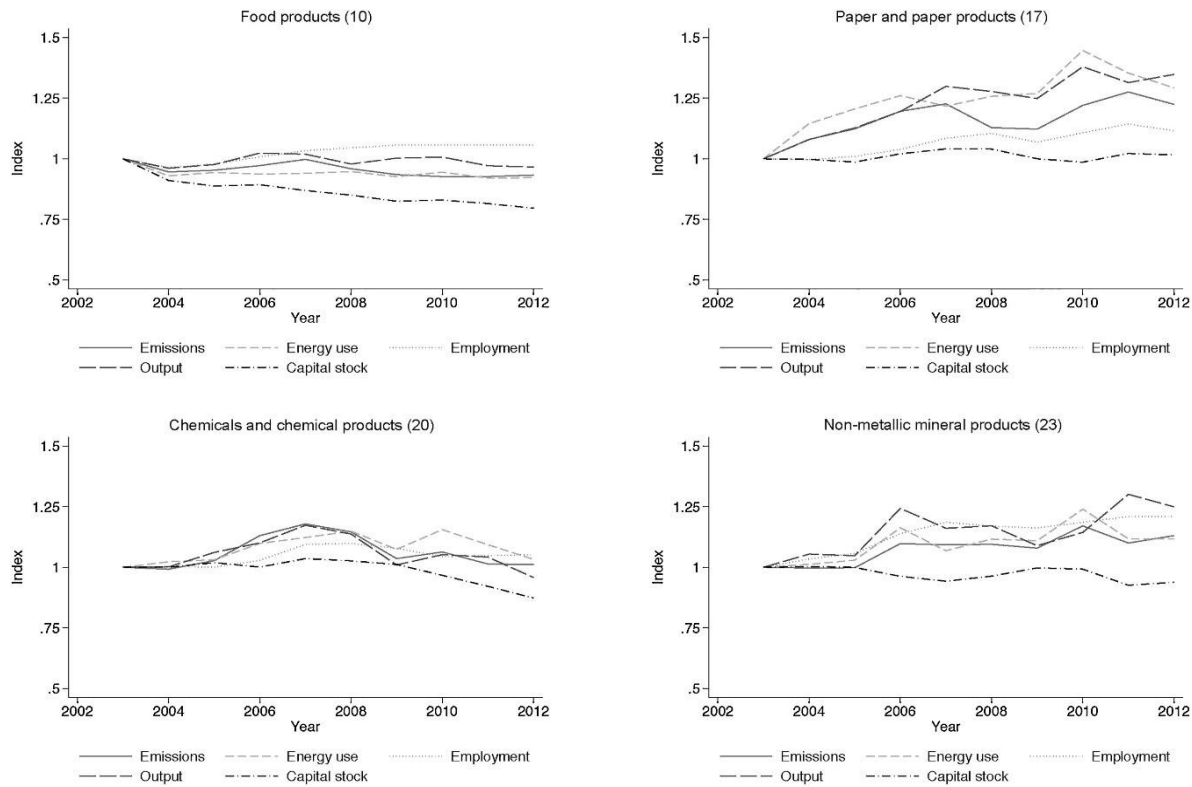

**Fig. 1.** Descriptive statistics: development across industries. *Notes:* Indexed medians (base year 2003) for emissions, energy use, employment, output, and capital stock. Source: Research Data Centres of the Statistical Offices Germany (2014): Official Firm Data for Germany (AFiD) – AFiD-Panel Industrial Units and AFiD-Module Use of Energy, own calculations.

We have plotted the development of the indexed median (base year 2003) of each variable for the industries manufacture of food products (10), manufacture of paper and paper products (17), manufacture of chemicals and chemical products (20), and manufacture of non-metallic mineral products (23), such as glass and cement. (which accounts for more than half of the German manufacturing firms regulated by the EU ETS)

### Inference:

While the development of output as well as input use in the food industry was barely affected by the economic crisis, the other graphs for these industries show a strong impact on output, emissions, and energy use in 2009.

This will be also reflected in the distances to the frontier, since firms produced less in the crisis year while they were not able to adjust their capital stock and their use of labour in the short term. The former can only be adjusted through investment or the disposal of physical capital while labour use also cannot be freely adjusted in Germany due to strong labour market regulation and collective labour agreements.

# 13. RESULTS

## 13.1 Stochastic production frontiers and efficiency

**Table 3** shows the parameter estimates of separate Cobb-Douglas production frontiers estimates for each two-digit industry

Table 3
Parameter estimates production frontier.

| Industry (NACE) | # Firms | Capital | Labor | Energy | Constant | $\hat{\sigma}_u$ | $\hat{\mu}_v$ | $\hat{\sigma}_v$ |
|---|---|---|---|---|---|---|---|---|
| Food products (10) | 6935 | 0.265 (0.010) | 0.323 (0.016) | 0.481 (0.014) | 2.047 (0.042) | 0.609 (0.010) | −443.728 (6.786) | 11.969 (0.384) |
| Beverages (11) | 703 | 0.223 (0.032) | 0.725 (0.050) | 0.257 (0.036) | 2.252 (0.188) | 0.549 (0.023) | −365.726 (119.853) | 9.839 (2.725) |
| Textiles (13) | 1103 | 0.199 (0.020) | 0.738 (0.037) | 0.117 (0.016) | 3.652 (0.104) | 0.507 (0.019) | −512.914 (19.645) | 13.732 (0.832) |
| Leather and related products (15) | 231 | 0.203 (0.050) | 0.742 (0.081) | 0.177 (0.045) | 3.308 (0.251) | 0.514 (0.041) | −908.894 (27.054) | 24.299 (1.631) |
| Wood and products of wood and cork (16) | 1587 | 0.186 (0.017) | 0.794 (0.029) | 0.146 (0.012) | 3.507 (0.079) | 0.498 (0.016) | −513.569 (16.601) | 13.775 (0.552) |
| Paper and paper products (17) | 1104 | 0.178 (0.021) | 0.677 (0.031) | 0.183 (0.012) | 3.720 (0.089) | 0.389 (0.017) | −360.647 (15.225) | 9.668 (0.581) |
| Printing and reproduction of recorded media (18) | 2255 | 0.115 (0.013) | 0.689 (0.026) | 0.250 (0.014) | 3.580 (0.061) | 0.367 (0.011) | −363.949 (73.232) | 9.821 (1.284) |
| Chemicals and chemical prducts (20) | 1722 | 0.205 (0.024) | 0.596 (0.029) | 0.173 (0.014) | 4.372 (0.092) | 0.522 (0.016) | −607.081 (40.582) | 16.373 (0.883) |
| Rubber and plastic products (22) | 3935 | 0.155 (0.011) | 0.726 (0.017) | 0.178 (0.010) | 3.645 (0.047) | 0.416 (0.008) | −385.313 (44.152) | 10.408 (0.750) |
| Other non-metallic mineral products (23) | 2446 | 0.206 (0.014) | 0.612 (0.020) | 0.111 (0.009) | 4.229 (0.070) | 0.501 (0.013) | −471.085 (5.073) | 12.644 (0.427) |
| Basic metals (24) | 1274 | 0.241 (0.024) | 0.637 (0.040) | 0.163 (0.019) | 3.617 (0.096) | 0.610 (0.019) | −300.333 (11.398) | 8.107 (0.761) |
| Fabricated metal products (25) | 9676 | 0.103 (0.006) | 0.896 (0.011) | 0.112 (0.006) | 3.791 (0.030) | 0.458 (0.006) | −372.983 (1.690) | 10.107 (0.190) |
| Electrical equipment (27) | 3077 | 0.170 (0.011) | 0.834 (0.021) | 0.071 (0.011) | 4.088 (0.049) | 0.449 (0.010) | −501.796 (6.310) | 13.482 (0.360) |
| Machinery and equipment n.e.c. (28) | 8620 | 0.071 (0.006) | 1.066 (0.011) | 0.027 (0.007) | 4.092 (0.032) | 0.453 (0.006) | −404.643 (2.223) | 10.965 (0.204) |
| Motor vehicles, trailers, and semi-trailers (29) | 1681 | 0.167 (0.017) | 0.893 (0.029) | 0.067 (0.019) | 3.840 (0.072) | 0.589 (0.020) | −405.271 (53.125) | 10.924 (1.072) |
| Furniture (31) | 1532 | 0.133 (0.013) | 1.034 (0.026) | 0.036 (0.016) | 3.599 (0.071) | 0.433 (0.014) | −409.503 (7.892) | 11.030 (0.481) |

Notes: The number of observations includes all firms that were active during the period from 2003 to 2012. We do not consider the industries manufacture of tobacco products (12), manufacture of wearing apparel (14), manufacture of pharmaceutical products (21), manufacture of computer, electronic and optical products (26), manufacture of other transport equipment (30), other manufacturing (32), and repair and installation of machinery and equipment (33). Source: Research Data Centres of the Statistical Offices Germany (2014): Official Firm Data for Germany (AFiD) – AFiD-Panel Industrial Units and AFiD-Module Use of Energy, own calculations.

We note that the estimated parameters of the stochastic production frontier vary across industries reflecting the strong heterogeneity within the manufacturing sector.

For the majority of industries, we observe statistically significant increasing economies of scale varying from 0.93 (manufacture of other non-metallic mineral products; 23) to 1.20 (manufacture of beverages; 11)

## 13.2 Figure 2 shows the distance to the production frontier of the median firm (for both treated and untreated) as a proxy for the development of efficiency over time

(This is plotted only by using differences in differences method along with semi-parametric assumptions without employing PSM NN matching techniques)

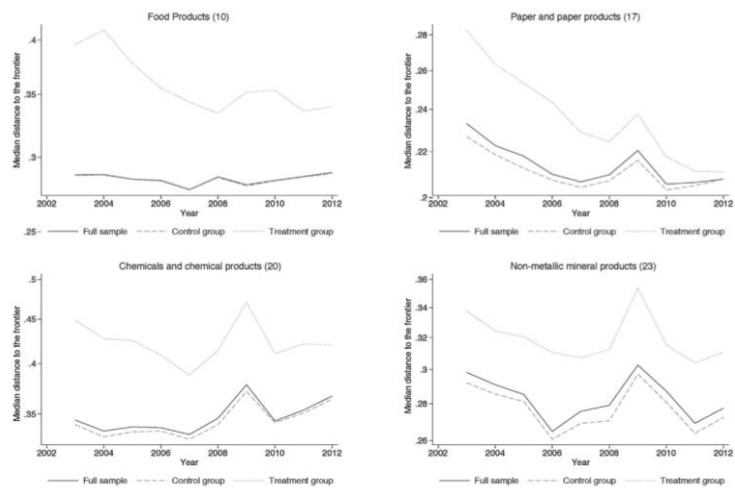

Fig. 2. Comparison treatment and control group: median distance to the production frontier. Notes: The vertical axis is displayed in log scale. Source: Research Data Centres of the Statistical Offices Germany (2014): Official Firm Data for Germany (AFiD) – AFiD-Panel Industrial Units and AFiD-Module Use of Energy, own calculations.

**SALIENT OBSERVATIONS:**

1.The dynamics of the distance to the production frontier reflect two developments:

a.  First, we observe that in all four industries, the median distance to the production frontier decreases during the early 2000s, i.e. the median firm becomes more efficient relative to the firms operating on the frontier. This trend in efficiency is driven by technological progress.

> Over time, the median distance to the production frontier decreases, *since technological progress gradually pushes the firms toward the frontier*.

b.  Secondly, we observe increases in the distance to the production frontier from 2006 onwards coinciding with the economic crisis. The distance to the production frontier peaks in 2009, the year when the crisis hit German manufacturing hardest. While demand and thus the production of goods rapidly decrease, firms do not adjust their capacity at the same speed. Therefore, low utilization rates increase the distance to the production frontier during the economic crisis

2. The distance to the production frontier of the median firm in the treatment group is higher than the distance to the production frontier of the median firm in the control group, indicating that the treated median firm operates less efficiently in these industries. Also, the closeness of lines indicate that the share of control firms is high. For the industries manufacture of food products (10), manufacture of paper and paper products (17), and manufacture of chemicals and chemical products (20) the distance between the treatment and control group decreases over time and converges toward the distance to the production frontier of the median firm in the control group. However, the industry of manufacture of non-metallic mineral products (23) does not show such a development.

## 13.3 IMPROVISING THE MODEL

Improvised results after implementing a combination of nearest neighbour matching with replacement i.e. unregulated firms can be used multiple times as a match.

Nearest neighbor matching treatment effects.

|  | One neighbor | Five neighbors | Twenty neighbors |
|---|---|---|---|
| *Year by year comparison (base year 2003)* | | | |
| 2005 | −0.0158 (0.0149) | −0.0343* (0.0134) | −0.0277* (0.0120) |
| 2006 | −0.0169 (0.0153) | −0.0121 (0.0133) | −0.0087 (0.0134) |
| 2007 | −0.0152 (0.0224) | 0.0015 (0.0167) | −0.0080 (0.0154) |
| 2008 | −0.0171 (0.0247) | −0.0001 (0.0172) | 0.0013 (0.0288) |
| 2009 | 0.0013 (0.0288) | 0.0069 (0.0222) | −0.0018 (0.0193) |
| 2010 | −0.0226 (0.0293) | −0.0038 (0.0228) | −0.0066 (0.0200) |
| 2011 | −0.0021 (0.0318) | −0.0129 (0.0295) | −0.0082 (0.0202) |
| 2012 | 0.0190 (0.0316) | 0.0029 (0.0334) | 0.0122 (0.0215) |
| *Comparison trading periods with pretreatment period* | | | |
| Phase I | −0.0289* (0.0124) | −0.0280* (0.0119) | −0.0265* (0.0108) |
| Phase II | −0.0294 (0.0222) | −0.0097 (0.0181) | −0.0164 (0.0166) |

Notes: Standard errors are robust with regard to heteroskedasticity and intra-firm correlation.
* Significant at the 5% level.
Source: Research Data Centres of the Statistical Offices Germany (2014): Official Firm Data for Germany (AFiD) – AFiD-Panel Industrial Units and AFiD-Module Use of Energy, own calculations.

The upper panel in above table shows estimated treatment effects for year by year comparisons (base year is 2003). The lower panel shows estimated treatment effects for Phase I and Phase II.

Pooling the data for the compliance periods, we find a significant negative effect of the EU ETS on firm specific distance to the production frontier during Phase I. The parameter estimates range between −2.65 and −2.89 percent. On average, the distance to the frontier decreased of manufacturing firms decreased during Phase I by −2.10 percent in comparison to the pre-treatment period. The parameter estimates for the treatment effect in Phase II are of the same magnitude but statistically insignificant.

## **13.4 SIGNIFICANT IMPACT ON PAPER INDUSTRY**

While implementing a parametric differences-in-differences model, heterogeneity across industries within the manufacturing sector might lead to insignificant treatment effects for the manufacturing sector as a whole. For example the industries manufacture of food products, chemicals and chemical products, and other non-metallic mineral products (cement, glass, etc.), we do not find a significant effect of the EU ETS on the distance to the production frontier, when controlling for confounding factors. However, For the paper industry, however, we find statistically and economically significant treatment effects for all specifications and time periods considered( including compliance phase 2)

Empirical evidence (shown below) that the EU ETS was an important driver of efficiency development among regulated firms in the paper industry.

The graphs explain themselves

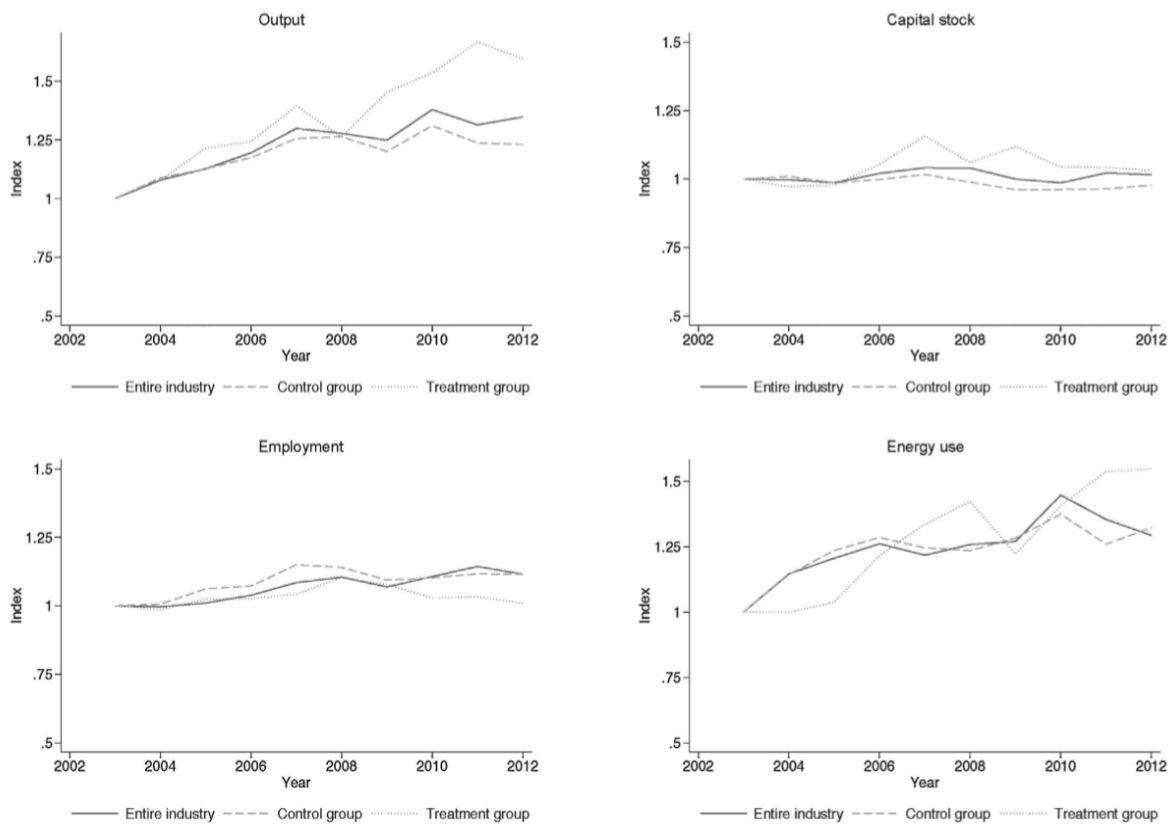

**Fig. 3.** Comparison treatment and control group – manufacture of paper and paper products. *Notes*: Source: Research Data Centres of the Statistical Offices Germany (2014): Official Firm Data for Germany (AFiD) – AFiD-Panel Industrial Units and AFiD-Module Use of Energy, own calculations.

# 14 CONCLUSION

The results based on the nonparametric nearest neighbour matching suggest a statistically significant positive effect of the EU ETS on the economic performance of the regulated firms during the Phase I of the EU ETS. A year-by-year analysis shows that the effect was only significant during the first year of Phase I. The EU ETS therefore had a particularly strong effect when it was introduced.

It is important to note that the EU ETS does not homogeneously affect firms in the manufacturing sector. We found a significant positive impact of EU ETS on the economic performance of regulated firms in the paper industry.

Although we cannot fully clarify the mechanisms at work, we conjecture that the EU ETS might have incentivized investments in more efficient capital stock that allowed the firms to produce more output with less inputs. Alternatively, the regulation by the EU ETS, might also have helped firms to overcome obstacles that impede in the short run the efficient use of existing capacities and in the long run process and product innovation.

*In conclusion, our results can be seen as an indication that the "strong" version of the Porter Hypothesis might apply to the case of the EU ETS – at least with regard to some specific industries.*

# 15 Observations from other papers analysing the impact of EU Ets on different countries

1. Wagner et al. (2014) show that the EU ETS reduced emissions of French manufacturing plants, by 15 to 20 percent on average between 2007 and 2010. They also found a significant decrease in employment in regulated plants of about 7 percent during the second compliance period of the EU ETS.

2. Jaraitė and Di Maria (2016) investigate the impact of the EU ETS on Lithuanian firms employing nearest-neighbour and kernel matching. They find that the EU ETS did not reduce $CO_2$ emissions, but improved $CO_2$ intensity. They do not find a significant effect on profits. However, regulated firms in Lithuania retired parts of their less efficient capital stock and made additional investments in the end of the second compliance period.

They do not reject the null hypothesis that the EU ETS had no causal impact on the profitability of Lithuanian firms

3. Klemetsen et al. (2016) use a parametric difference-in-differences approach in order to isolate and quantify the effect of the EU ETS on emissions, emission intensity, value added, and labour productivity of Norwegian plants. They find that the EU ETS decreased emissions and at the same time *increased value added and labour productivity during the second compliance period.*

They show that the effect of the EU ETS on value added and labour productivity of regulated firms in Norway was **positive**.

4. Calel and Dechezlepretre (2016) examine the effect of the EU ETS on technological change, in particular patenting. They combine patent and commercial firm-level data for Europe with data from the EU ETS. Using a matching approach, they find that the EU ETS increased the number of low-carbon patents among regulated firms by 10 percent between 2005 and 2010 while not crowding out patenting for other technologies

5. Lutz (2016) estimates a structural production function that allows for endogenous productivity and employs a parametric difference-in-differences approach in order to quantify the effect of the EU ETS on firm-level productivity. He measured productivity as a deviation in the mean production function levels.

*He shows that the EU ETS had a significant positive impact on productivity during the first compliance period*

*On the whole, our research models' conclusions converge with the other models and clearly indicate that although EU ETS reduced carbon emissions it had no statistically significant negative impact on competitiveness, productivity and employment of regulated firms.*

*On the contrary some show a causal statistically **significant positive impact**.*

*These claims provide a robust empirical evidence in support of the "strong" version of the Porter hypothesis **that EU ETS indeed induced technological change and also led to an effective utilization of under-utilized resources.***

# 16 SCOPE OF SUCH AN EMISSION TRADING SCHEME IN INDIA: PRECAUTIONARY MEASURES

1. **Preventing overallocation of tradable units under the Emission Scheme**

In the case of the EU, a tremendous over-allocation of free EUAs during Phase I led to a decline in EUA prices from above EUR 25 to zero in 2007.

In Phase II, the EU ETS once again suffered from massive over-allocation, but this time it was due to the economic-crisis. This happened because the fear of economic crisis caused the regulatory bodies to over-supply the energy usage to boost productivity. This unadjusted supply of free allowances led to an oversupply of allowances and as a result these developments, the EUA price decreased from more than EUR 25 at the beginning of Phase II to less than EUR 10 in the second half of Phase II.

So, from the beginning of Phase III the EU modified their approach by decentralizing the allocation of EUA's and gradually shifting from grandfathering to auctioning.

Following the "learning-by-doing hypothesis", we should first cautiously allocate these tradable units so that there is no oversupply in the market. This will compel the companies to cost-effectively use their existing allocated units. This could be done by improving heat recovery and bringing on process innovations.

2. **Preparing before-hand for the eventualities such as**
a. the initial dip in employment level
b. decline in production levels
c. decreasing export levels due to an increase in input cost

This could be effectively tackled if the government takes the following steps:
1. Providing subsidies in renewable resources. For eg: Solar cells, wind turbines, hydro energy etc

2. Increasing the investment in human capital by training the unskilled workers in future technologies
3. More investment in R&D to boost innovation
4. India is a huge importing country, so regulating Indian manufacturing firms will result in less of global competition and more of local competition. This will in turn boost competitiveness and lead to an increase in the overall export

All these measures will curb extra costs due to ETS and prevent the decrease in export levels.

# References

1.Andreas Löschel et al.,The impacts of the EU ETS on efficiency and economic performance – An empirical analyses for German manufacturing firms  Andreas Löschel,Benjamin Johannes Lutz,Shunsuke Managi, Resource and Energy Economics,  May 2019

2.Calel and Dechezlepretre,Environmental policy and directed technological change: evidence from the European carbon market Rev. Econ. Stat. (2016)

3.Ellerman et al., 2010 Pricing Carbon: The European Union Emissions Trading Scheme,Cambridge University Press, United Kingdom (2010)

4.Klemetsen et al., 2016 The Impacts of the EU ETS on Norwegian Plants' Environmental and Economic Performance, NMBU Working Papers (2016)

5.Jaraitė and Di Maria, 2016, Did the EU ETS make a difference? An empirical assessment using Lithuanian firm-level data,Energy J., 37 (1) (2016), pp. 1-23

6.Lutz, 2016 Emissions Trading and Productivity: Firm-level Evidence from German Manufacturing, ZEW Discussion Paper No. 15-013 (2016)

7.Martin et al., 2016 The impact of the European Union Emissions Trading Scheme on regulated firms: what is the evidence after ten years? Rev. Environ. Econ. Policy, 10 (1 (January)) (2016), pp. 129-148

8.Petrick and Wagner, 2014 The Impact of Carbon Trading on Industry: Evidence from German Manufacturing Firms (2014)